\documentclass[aps,preprint,amssymb,amsmath,nofootinbib]{revtex4}
 
 \newtheorem{theorem}{Theorem}[section]
 \newtheorem{trial definition}{Trial definition}[section]

 \newenvironment{hypothesis}{\medskip HP:} {\medskip}
 \newenvironment{thesis}{\medskip TH:} {\medskip}
 \newenvironment{proof}{\begin{center}PROOF: \end{center}} {
$ \blacksquare $}

\begin{document}
 \title{A remark about the Mermin-Squires Music Hall's inteludium}
 \author{Gavriel Segre}
 \homepage{http://www.gavrielsegre.com}
 \email{info@gavrielsegre.com}
 \date{12-3-2004}
 \begin{abstract}
 The Mermin-Squires Music Hall inteludium on the Einstein-Podolsky-Rosen affair is
 analyzed by showing the fallacity of the One-Borel-Normality
 Criterion and the necessity of replacing it with the more
 restrictive Algorithmic-Randomness Criterion
\end{abstract}
\maketitle
\newpage
Let us consider the Squires' Music Hall's inteludium
\cite{Squires-94} based on Mermin's analysis of the
Einstein-Podolski-Rosen experiment \cite{Mermin-90}:

\smallskip

Alice and Bob are able of executing a show at first sight
astonishing:

they are closed in two separated boxes of a theatre's stage so
that they can't communicate each other.

At regular intervals two public's people, let's say one from the
right side and the other from the left side of the parterre, give
respectively to Alice and Bob a ticket on which it is written one
of the numbers 1, 2 or 3.

The choice of the numbers appearing on the tickets the two
spectators take to the artists is absolutely random.

At this point Alice and Bob are asked to write YES or NO on the
received ticket.

Obviously Alice can't know which number is written on Bob's
ticket and viceversa.

\smallskip

The show's exceptionality is that \textbf{all the times Alice and
Bob receive a couple of tickets on which it is  written the same
number, they answer in the same way}

\medskip

Let us now suppose to be in the public and to look for a
criterion in order to establish if Alice and Bob are really
telepathic or they are simply swindlers.

\smallskip

Indeed anyone would think that there is a very simplest
explanation of Alice and Bob's performance: \textbf{they agreed
before the show on how to answer}.

\smallskip

Is there, then, a way of certifying if this is the case?

\smallskip

Let us adopt the following notation:
\begin{itemize}
  \item $ t_{A} (n) \, , t_{B} (n) $ denote the
tickets given, respectively, to Alice and Bob at the $ n^{th} $
repetition of the performance
  \item $  a_{A} (n) \, , a_{B} (n) $ denote the
answers  written, respectively, by Alice and Bob on the $ n^{th}
$ ticket
\end{itemize}

\smallskip

By hypothesis $ \{ t_{A} (n) \} $ and $ \{ t_{B} (n) \} $ are two
independent sequence of i.i.d.  uniform random variables over $ \{
1 , 2 , 3 \} $:
\begin{align} \label{eq:distribution of tickets}
  Prob( t_{A} (n)  \; & = \; i ) \; =  \; \frac{1}{3} \; \; \forall i \in \{
1 , 2 , 3 \} , \forall n \in {\mathbb{N}}  \\
 Prob( t_{B} (n)  \; & = \; i ) \; =  \; \frac{1}{3} \; \; \forall i \in \{
1 , 2 , 3 \} , \forall n \in {\mathbb{N}}
\end{align}

\smallskip

Let us then introduce the \textbf{concordance sequence} $ \bar{C}
\in \{ 0 , 1 \}^{\infty} $ whose $ n^{th} $ digit $ C_{n} $ is
defined as:
\begin{equation}
 C_{n} \; := \;
  \begin{cases}
    1 & \text{if $ a_{A} (n) \, = \, a_{B} (n) $ }, \\
    0 & \text{otherwise}.
  \end{cases}
\end{equation}

\medskip

Squires introduced the following:

\begin{theorem}
\end{theorem}
CRITERION OF 1-BOREL NORMALITY

\begin{hypothesis}
\end{hypothesis}
\begin{multline}
  \exists \, \phi \,  \in \, RECURSIVE-MAPS ( \{ 1 , 2 , 3 \} , \{ YES
, NO  \} ) \, : \\
\, a_{A} (n) = \phi ( t_{A} (n) ) \; and \;
a_{B} (n) = \phi ( t_{B} (n) )
\end{multline}
\begin{thesis}
\end{thesis}
\begin{equation}
   p_{1-Borel} ( \bar{C} ) \; \; \text{ doesn't hold}
\end{equation}
\begin{proof}
We have clearly that:
\begin{equation}
  N_{0} ( \phi(1) \phi(2) \phi(3) ) \; \neq \;  N_{1} ( \phi(1) \phi(2) \phi(3) )
\end{equation}
By the eq.\ref{eq:distribution of tickets} and the Law of Large
Numbers it immediately follows that $ \bar{C} $ can't be 1-normal.
\end{proof}

\medskip

Anyway Squires doesn't consider the case in which Alice and Bob
adopt a more clever way of cheating, i.e. they use a previously
concorded answering-algorithm depending also from n; in this way
they may easily to scoff at Squires' criterion of
Borel-1-normality.

\smallskip

Let us suppose, for example, that each of them answers according
to the following rule:
\begin{equation}
  a(n) ( t ) \; := \;
  \begin{cases}
    YES & \text{if $ t = 1 $ }, \\
    NO & \text{if $ t = 2 $ }, \\
    YES & \text{if $ t = 3 $ and n is even}, \\
    NO & \text{if $ t = 3 $ and n is odd},
  \end{cases}
\end{equation}

\smallskip

Since in this case the according-rule doesn't break the balance
among YES and NO, by the eq.\ref{eq:distribution of tickets} and
the Law of Large Numbers it follows immediately that $ \bar{C} $
is Borel-1-normal.

\medskip

So if Alice and Bob answer according to this rule, they scoff at
Squires according to which they appear as really telepathic.

\smallskip

There exist, anyway, a way to unmask the deception; this, anyway,
requires the adoption of a stronger criterion:

\medskip

\begin{theorem}
\end{theorem}
CRITERION OF ALGORITHMIC RANDOMNESS

\begin{hypothesis}
\end{hypothesis}
\begin{multline}
  \exists \, \phi \,  \in \, RECURSIVE-MAPS ( \{ 1 , 2 , 3 \} \times {\mathbb{N} } , \{ YES
, NO  \} ) \, : \\ a_{A} (n) = \phi ( t_{A} (n) \, , \, n ) \; and
\; a_{B} (n) = \phi ( t_{B} (n) \, , \, n  )
\end{multline}
\begin{thesis}
\end{thesis}
\begin{equation}
  \bar{C} \; \notin \; RANDOM( \{0,1\}^{\infty} )
\end{equation}
\begin{proof}
The Chaitin-Schnorr Theorem \cite{Calude-02}  states  the
equivalence of Martin-L\"{o}f statistical characterization of
algorithmic-randomness and Chaitin's one as algorithmic
incompressibility:
\begin{equation}
  \bar{x} \in RANDOM( \{0,1\}^{\infty} ) \; \Leftrightarrow \;
  \exists c > 0 \, : \, I ( \vec{x} (n) ) \, \geq \, n - c \,
  \forall n \geq 1
\end{equation}
where:
\begin{equation}
  I ( \vec{x} ) \; := \; \min \{ | \vec{u} | \, : \, U ( \vec{u} , \lambda  ) = \vec{x}  \}
\end{equation}
and where U is a fixed \textbf{Chaitin universal computer}.

\smallskip

But by  hypothesis Alice and Bob's answer is
algorithmically-compressible through the rule $  \phi \,  \in \,
RECURSIVE-MAPS ( \{ 1 , 2 , 3 \} \times {\mathbb{N} } , \{ YES ,
NO \} ) $ they concorded before the show, implying the thesis
\end{proof}

\end{document}